\documentclass{ptapap}

\author{Konstanze Zwintz}[UIBK]
\author{the BRITE team}
\affil[UIBK]{Universit\"at Innsbruck, Institut f\"ur Astro- und Teilchenphysik\\
  Technikerstrasse 25/8, A-6020 Innsbruck}

\title{The exoplanet host star $\beta$ Pictoris seen by BRITE-Constellation\thanks{Based on data collected by the BRITE Constellation satellite mission, designed, built, launched, operated and supported by the Austrian Research Promotion Agency (FFG), the University of Vienna, the Technical University of Graz, the Canadian Space Agency (CSA), the University of Toronto Institute for Aerospace Studies (UTIAS), the Foundation for Polish Science \& Technology (FNiTP MNiSW), and National Science Centre (NCN).}}

\begin{document}

\maketitle

\begin{abstract}

BRITE-Constellation has observed the exoplanet host star $\beta$ Pictoris for 225 days from November 2016 to June 2017. These data allow for an accurate description of the pulsational properties and an asteroseismic interpretation. They were also observed as part of an international observing campaign which aims to detect the transit of $\beta$ Pictoris b's Hill sphere and study the circumstellar disk around $\beta$ Pictoris itself.

\end{abstract}

\section{Introduction}

$\beta$ Pictoris is the only $\delta$-Scuti like pulsating \citep{koen2003phot,koen2003spec} pre-main sequence or very early ZAMS star accessible to BRITE due to its brightness of $V=3.86$ mag. 

The $\delta$ Scuti like pulsations in $\beta$ Pictoris were first discovered by \citet{koen2003phot} using photometric measurements obtained with the SAAO 0.5 and 0.75m telescopes and photomultipliers with neutral density filters. Using data obtained in four observing nights, \citet{koen2003phot} found three pulsation frequencies at 47.055\,d$^{\rm -1}$, 38.081\,d$^{\rm -1}$ and 52.724\,d$^{\rm -1}$ with amplitudes of 1.63, 1.50 and 1.07 mmag respectively. 
On the basis of this discovery,  \citet{koen2003spec} obtained 697 high-dispersion spectra with GIRAFFE at the SAAO 1.7m telescope over a period of two weeks and found 18 pulsation frequencies in the spectroscopic line profiles.

The $\beta$ Pictoris system also includes a wide, dense circumstellar disk that is seen edge-on and a giant gas planet ($\beta$ Pictoris b) that was imaged in the L' band (3.78 microns) using VLT / NaCo \citep{lagrange2010}. As the inclination of the planet is 88.81$^{\circ}$ $\pm$ 0.12$^{\circ}$ as seen from Earth \citep{wang2016}, $\beta$ Pictoris b will not transit its host star. But according to the calculated orbital parameters, the Hill sphere of $\beta$ Pictoris b is expected to transit the star with a closest approach either in June 2017, August 2017 \citep{wang2016}, October 2017 (A. M. Lagrange, private communication) or up to January 2018 \citep{lecavelier2016}.

\section{The $\beta$ Pictoris observational campaign}
As the asteroseismic interpretation in 2003 did not distinguish between the pre-main sequence or early main sequence evolutionary stage of $\beta$ Pictoris and the photometric time series obtained from ground are very short \citep{koen2003phot}, $\beta$ Pictoris was proposed for BRITE-Constellation observations already in 2015. Hence, an observing run with the BRITE-Constellation satellites BRITE-Toronto (BTr) and BRITE Lem (BLb) was scheduled for the period between November 2016 and April 2017.

The predicted transit of the Hill sphere of $\beta$ Pictoris b triggered an international observing campaign involving several facilities on the ground and in space. It included dedicated windows of observations with the Hubble Space Telescope (HST), continuous photometric time series with the bRing telescope in South Africa \citep{bRing2017} and the ASTEP 400 telescope installed at Concordia station, Dome C, in Antarctica (Mekarnia et al. 2017, submitted), and spectroscopic observations using ESO HARPS (A. M. Lagrange, PI). BRITE-Constellation data are and will be used for the investigation of the transit of the Hill sphere and the disk surrounding $\beta$ Pic.

\subsection{BRITE-Constellation}
\begin{figure}
   \begin{center}
   \includegraphics[width=\textwidth]{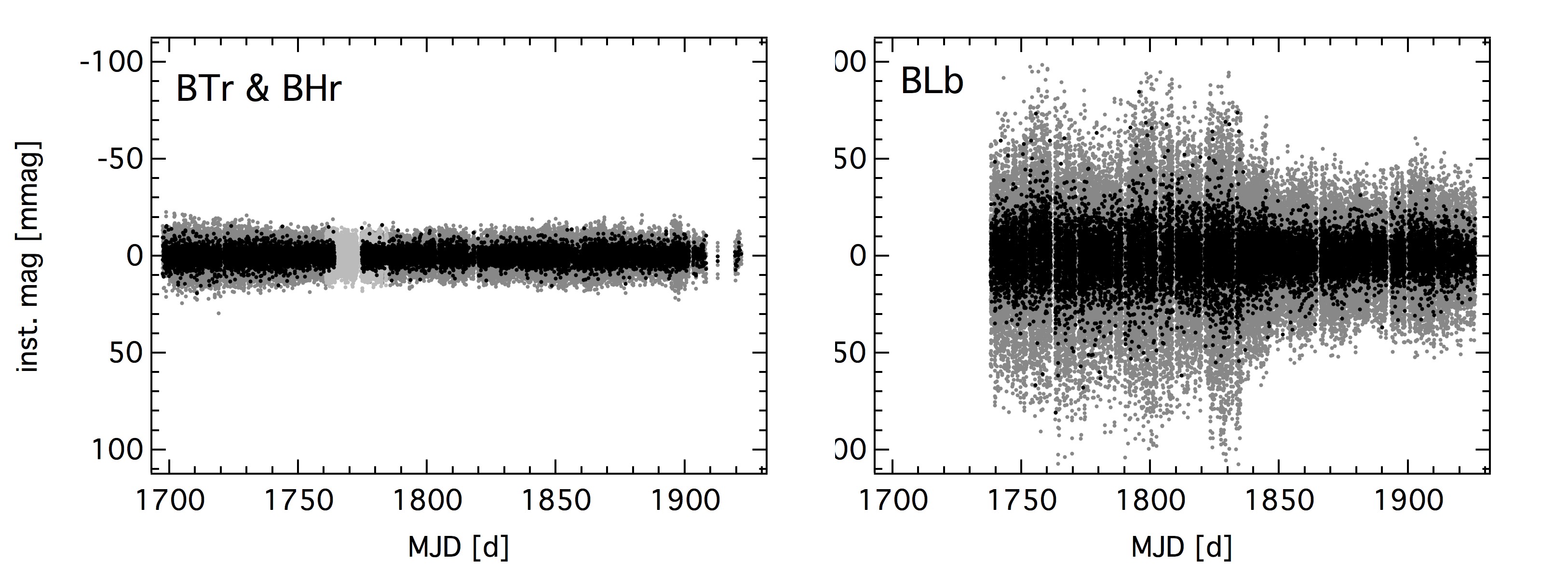}
   \caption{Light curves of $\beta$ Pictoris obtained by BTr and BHr (left panel) and BLb (right panel). Grey symbols are the original data and black symbols are data binned to 2-minute intervals. The light grey symbols in the left panel correspond to the BHr data set, the dark grey symbols to the BTr data.}
   \label{lcs}
   \end{center}
   \end{figure}

The $\beta$ Pictoris data were obtained during the BRITE-Constellation observing run 23-VelPic-II-2016. 
BRITE-Toronto (BTr) obtained observations of $\beta$ Pictoris in the red filter from 4 November 2016 to 22 June 2017 for in total 225 days, while BRITE-Lem (BLb) observed for 188 days continuously in the same period of time (see Figure \ref{lcs}). There is a small gap in the BTr data set during which no observations could be taken with this satellite. During this time, the BRITE-Heweliusz (BHr) satellite took over the observations of the VelPic-II-2016 field providing the needed continuous time coverage in the red filter. 
The observations also covered the first predicted time for the transit of the Hill sphere on 16 June 2017 where the BRITE-Constellation data were also used to confirm that no decrease in intensities caused by the transit was observed.

The BRITE observations were conducted in chopping mode where the position of the target star within the CCD plane is constantly alternated between two positions about 20 pixels apart on the CCD. This procedure was adopted to mitigate the impact of high dark current on the CCDs \citep[see ][]{popowicz2017}.

\section{Data analysis}
The BRITE-Constellation data of $\beta$ Pictoris were corrected for instrumental effects according to the iterative procedure described by \citet{pigulski2016} and modified to include two-dimensional decorrelations. 
The whole procedure included outlier rejection, 1D and 2D decorrelations, and removal of corrupted orbits (e.g. affected by poor stability of the satellite).
The decorrelation was made sequentially, allowing for multiple decorrelations with the same parameter (or a pair of parameters).
At each step, the parameter showing the strongest correlation was chosen for correction. The strength of a correlation was defined as the degree of reduction in variance due to decorrelation with a given parameter (for 1D) or a pair of parameters (for 2D).
The iterations were stopped when all correlations (both 1D and 2D) resulted in a variance reduction smaller than 0.05 per cent.
The whole procedure was run independently for each setup. Outliers and the worst orbits were rejected at least twice during the
whole procedure. After decorrelating the data, the blue and red data were separately combined, taking the mean magnitude offsets between the setups into account.

The complete BTr light curve of $\beta$ Pictoris consists of 53620 data points (see Figure \ref{lcs}, left panel), the complete BLb light curve of 74306 data points (see Figure \ref{lcs}, right panel). The short BHr data set that covers the gap in the BTr observations has 13958 data points; hence the complete BRITE light curve in the red filter includes 67578 data points. 

The frequency analysis of the BRITE photometric time series was performed using an iterative pre-whitening method based on the Lomb-Scargle periodogram, which is described by Van Reeth et al. (2015a). Frequencies are identified to be significant if they exceed 3.9 times the local noise level in the Fourier domain. Frequency errors are calculated using the method by Schwarzenberg-Czerny (2003), which is based on the statistical errors resulting from a non-linear least-squares fit corrected for the correlated nature of the data.

I total we identified 12 pulsation frequencies in the red filter and 3 frequencies in the blue filter, both in the range from 5 to 55\,d$^{\rm -1}$. The residual noise level is 47ppm in the BTr and 170ppm in the BHr data calculated from 0 to 100\,d$^{\rm -1}$.

   \begin{figure}
   \begin{center}
   \includegraphics[width=\textwidth]{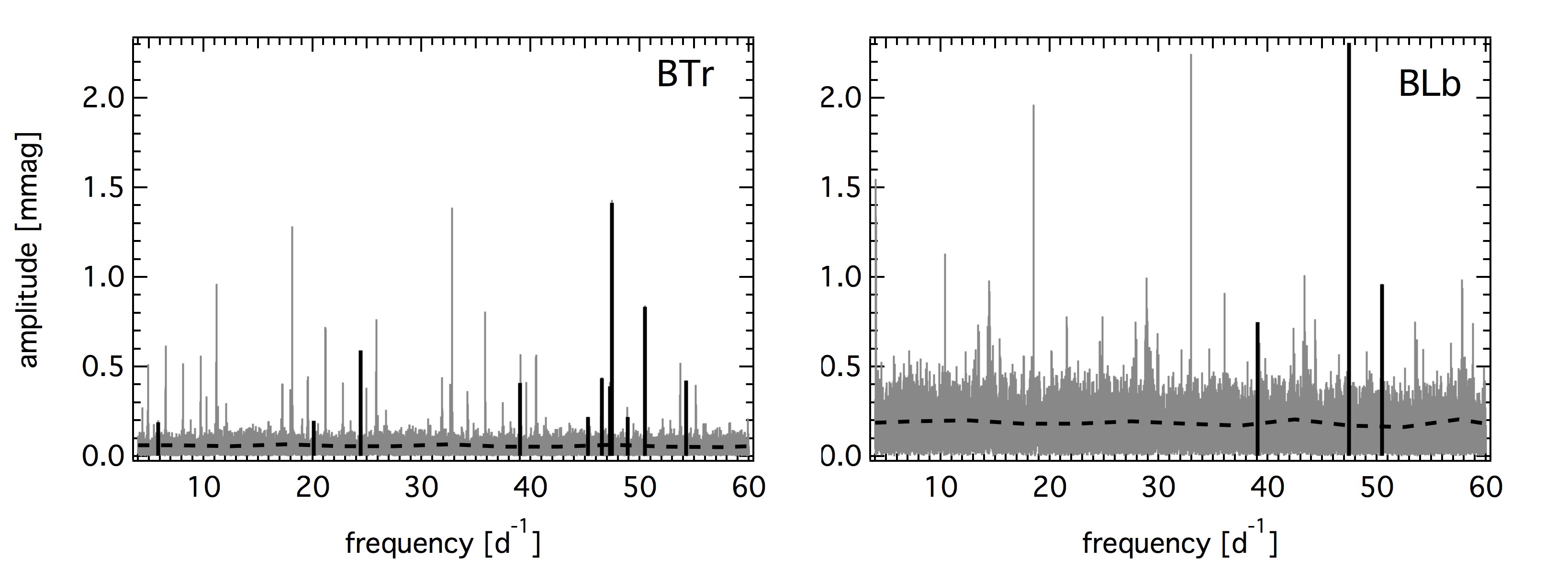}
   \caption{Amplitude specta of $\beta$ Pictoris obtained by BTr (left panel) and BLb (right panel) where the identified pulsation frequencies are marked in black and the Fourier noise spectrum is given as a black dashed line.}
   \label{amps}
   \end{center}
   \end{figure}

\section{Summary}
Using BRITE-Constellation data of $\beta$ Pictoris obtained with the BTr, BHr and BLb satellites in the field 23-VelPic-II-2016, we identified 12 pulsation frequencies of $\delta$ Scuti type. The ongoing asteroseismic interpretation will allow to constrain the star's interior structure and yield an asteroseismic estimate of $\beta$ Pictoris' age.

New BRITE-Constellation observations of $\beta$ Pictoris are currently scheduled for the period November 2017 to April 2018 in the field 33-VelPic-III-2017 with BHr. These observations will be in particular important for the predicted transit of the Hill sphere that so far has not been observed from any of the monitoring facilities. But they will also allow us to extend the total time base for the investigation of the pulsational content of this exoplanet host star.

\acknowledgements{Based on data collected by the BRITE Constellation satellite mission, designed, built, launched, operated and supported by the Austrian Research Promotion Agency (FFG), the University of Vienna, the Technical University of Graz, the Canadian Space Agency (CSA), the University of Toronto Institute for Aerospace Studies (UTIAS), the Foundation for Polish Science \& Technology (FNiTP MNiSW), and National Science Centre (NCN). \newline
KZ acknowledges support by the Austrian Fonds zur F\"orderung der wissenschaftlichen Forschung (FWF, project V431-NBL), the Austrian Space Application Programme (ASAP) of the Austrian Research Promotion Agency (FFG) and the TirolerWissenschaftsfonds (GZ: UNI-0404/1985;
PI: K. Zwintz).}

\bibliographystyle{ptapap}
\bibliography{kzwintz_betaPic}

\end{document}